\documentclass[12pt]{article}
\usepackage{graphicx}
\usepackage{natbib}
\bibpunct{(}{)}{;}{a}{}{,}
\usepackage{txfonts}
\def\p0943{PSR B0943$+$10}
\def\q0031{PSR B0031$-$07}
\begin{document}

\title{The issue of aliasing in \p0943 \\ II Signal processing arguments}
\author{M. Vivekanand\footnote{vivek@ncra.tifr.res.in} \\
	National Center for Radio Astrophysics, TIFR, \\
	Pune University Campus, P. O. Box 3, \\
	Ganeshkhind, Pune 411007, India.}
\maketitle

\noindent {\bf Abstract}:
\cite{DR1999, DR2001} claim that the frequency of the very narrow feature, in the 
spectrum of radio flux variations of \p0943, is an alias of its actual value. They 
also claim to have detected an amplitude modulation on the above phase modulation. 
This paper argues that both these claims are unjustified. 

\section{Introduction}

The rotation powered radio pulsar \p0943 exhibits sub pulses that are drifting 
systematically from period to period within the observable pulse window. This pulsar has 
a very narrow feature in the longitude resolved spectrum of intensity fluctuations; its 
$Q$, defined as its central frequency divided by its width, is relatively high 
(\citealt{TH1971}; \citealt{BRC1975}; \citealt{SO1975}). Recently 
\citeauthor{DR1999} (\citeyear{DR1999}, \citeyear{DR2001}; henceforth DR1999 and DR2001)
put its $Q$ at $\ge$ 500. It has been speculated that this very narrow spectral feature, 
occurring at 0.465 cycles per pulsar period [cpp], could be an alias of the actual value 
0.535 [cpp] (\citealt{SO1975}). \cite{DB1973} noted such a general possibility in radio 
pulsars and claimed, based on the phase information in the fluctuation spectrum, that PSR 
B2303+30 has an aliased spectral feature. This was contested by \cite{SO1975}, who state 
on their page 326 that ``even a phase analysis, contrary to Backer's statements 
(\citealt{DB1973}), is unable to decide between the two possibilities'', viz., whether 
the spectral feature is aliased or not.  Indeed, in a later paper Backer did not repeat 
such a claim for \p0943, whose fluctuation spectrum is similar to that of PSR B2303+30 
(\citealt{BRC1975}); he quoted the the true value as 0.465 [cpp]. However, DR1999 and 
DR2001 recently claimed that the spectral feature is indeed an alias; that the actual 
value is 0.535 [cpp]. In their view \cite{SO1975} ``came to the wrong conclusion'' 
(DR2001). They also claim that the weak, symmetrically spaced sidebands, at 0.027 [cpp] 
away from the above spectral feature ``strongly suggest ... a regular, highly periodic
amplitude modulation of the ... drifting sub pulse sequences''.

This paper argues that DR1999 and DR2001 are unjustified in drawing these two conclusions. 
The question, whether \p0943 has an aliased spectral feature or not, is still unresolved; 
and the latter observation is as likely, if not more likely,  due to an additional phase 
modulation of the drifting sub pulses.

\section{A review of the relevant signal processing}

The signal from an ideal pulsar with drifting sub pulses falls under the topic ``pulse 
position modulation'' (PPM); it consists of periodically occurring narrow pulses whose 
positions are modulated by another periodicity. Its general principles can be found in 
books on electronic communication engineering (see \citealt{BPL1998}). The spectrum of 
a PPM signal when the position modulation is due to a pure ``tone'' (a single frequency) 
is given by \cite{PP1965} on his page 541 and by \citeauthor{SBS1966} (\citeyear{SBS1966}; 
henceforth SBS1966) on their pages 252 -- 253.

\subsection{Frequency domain discussion}

To begin with let us assume that the drifting sub pulse pattern is PPM due to a pure 
tone. 

The abscissa in fig.~6-2-3 on page 251 of SBS1966 represents the phase of the sampling 
signal at which a pulse is observed; in our case the sampling signal has the pulsar 
period $P$ (sec), or frequency $1 / P$ Hz or 1 [cpp]. The ordinate represents the 
corresponding phase of the modulating signal; in our case it has period $P3$ pulsar 
periods (frequency $1 / P3$ [cpp]) which is the repetition time of the drifting sub 
pulse pattern. The motion in time of the drifting sub pulses in this figure is along 
a straight line of slope $1 / P3$. In the last para on their page 254, SBS1966 state 
``the average pulse repetition frequency should be at least twice the highest signal 
frequency in order to obtain the minimum number of samples necessary for satisfactory 
signal recovery''; i.e., $1 > 2 / P3$ ($\Rightarrow 1 / P3 < 0.5$) to avoid aliasing. 
This can be verified by considering two straight lines of slopes $< 0.5$ and $> 0.5$ 
in fig.~6-2-3 of SBS1966.

Thus, the Nyquist sampling criterion is the same for a PPM signal and a canonical pulse 
amplitude modulation (PAM) signal (amplitude modulation of periodic pulses). DR2001 are 
wrong when they claim in their conclusion that ``A harmonic resolved fluctuation 
spectrum uses the information within the finite width of the pulse to achieve a Nyquist 
frequency of 1 [cpp], showing clearly that the primary feature is aliased''. Their 
average pulse repetition frequency is obviously $P$ (sec); so their Nyquist frequency 
is only 0.5 [cpp]. Consequently, they are also wrong in concluding that ``the primary 
feature is aliased'' based merely upon the Fourier technique; they require {\bf 
additional and independent information}, as discussed ahead.

Consider a PAM signal in which the amplitude modulation is due to a pure tone of 
frequency $\nu$ [cpp]. The amplitudes occur at frequencies $\nu$ [cpp], $1 \pm \nu$ 
[cpp], $2 \pm \nu$ [cpp], etc; and there is no difference in the amplitude spectra of 
the original ($\nu < 0.5$) and the aliased ($\nu > 0.5$) signal. However the phase 
spectra differ. Therefore one can distinguish between the original and aliased PAM 
signals only by using additional information such as the phase of the modulating signal.
In the PPM case, the amplitudes in the spectra occur at frequencies $\nu$ [cpp], $1 \pm 
m \times \nu$ [cpp], $2 \pm m \times \nu$ [cpp], etc., where $m = 1, 2, 3, ...$ is the 
order of the harmonic (see eq. 6-2-13 on page 252 of SBS1966). Now, both amplitude and 
phase spectra differ for the original and aliased signals. Therefore once again, one 
can distinguish between the original and aliased PPM signals only by using additional 
information such as the exact shape of the two amplitude spectra. In practice it is 
impossible to predict this exactly in the current pulsar context.

\begin{figure}
\includegraphics[width=12.5cm]{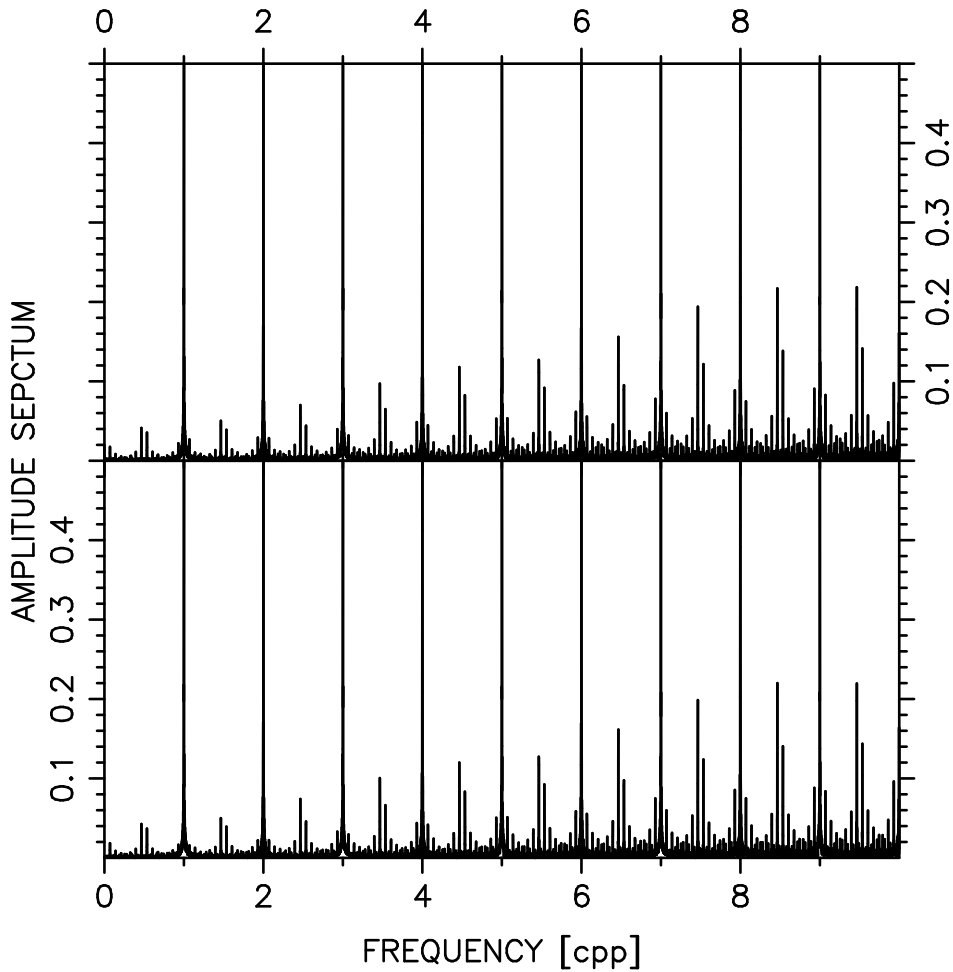}
\caption{
	Simulated amplitude spectra of a pulsar signal showing the drifting sub pulse 
	phenomenon. Only a small range of frequency has been shown for better visual 
	comparison. The time series consists of $4 \times 1024 \times 1024$ samples, 
	each of duration 0.25 milli seconds (ms). The pulsar parameters are those of 
	\p0943, taken from DR2001: period $P = 1.0977$ (sec), Gaussian integrated 
	profile of width $= 0.031 \times P$ (sec), maximum time departure of drifting 
	sub pulses $= 0.021 \times P$ (sec). The sub pulses are assumed to be Gaussian
	in shape of very narrow intrinsic width, of $\approx$ one sampling interval; 
	making the width a more realistic value merely suppresses the spectra at higher 
	frequencies. The drifting sub pulse phenomenon is modeled as a PPM signal with 
	a saw-tooth modulation in time. {\bf Top panel}: Original modulating frequency 
	$= 0.465$ [cpp], or $P3 = 2.1505376$ periods; {\bf Bottom panel}: aliased 
	modulating frequency $= 0.535$ [cpp], or $P3 = 1.8691589$ periods, and with 
	opposite drift direction.
	}
\label{fig1}
\end{figure}

Fig.~\ref{fig1} simulates the amplitude spectra of a pulsar signal showing the 
drifting sub pulse phenomenon, modeled as a PPM signal with a saw-tooth modulation 
in time, for both the original modulating frequency (top panel), and its alias but 
with the opposite drift direction (bottom panel). The pulsar parameters are taken 
from DR2001 for \p0943; however the figure is insensitive to small variations in 
these parameters. The algorithm was checked by reproducing the spectrum in eq. 6-2-13 
of SBS1966. The difference between the two amplitude spectra in fig.~\ref{fig1} is 
generally $\le$ 5\% of the maximum amplitude of pulsar harmonics, upto a frequency 
of 100 [cpp], for the harmonics $m = 1$ to $3$. For example, at $59.93$ [cpp], the 
second harmonics differ by $\approx$ 12.5\% in the two panels; but their average 
amplitude is $\le 33$\% of the maximum amplitude. The third harmonics, aliased to 
$59.605$ [cpp], differ by $\approx$ 9\%; but their average amplitude is only about 
4.5\% of the maximum amplitude. At $110.395$ [cpp], the harmonics differ by 
$\approx$ 20\%; but their average amplitude is now about 37.5\% of the peak amplitude 
of the pulsar harmonics; this is also insignificant.

Therefore at lower frequencies ($< 100$ [cpp]; the range analyzed by DR2001) the 
spectra of the original and the aliased signals should be predicted to an accuracy of 
$\le$ 5\% to tell the difference. This is almost impossible, considering other factors 
ignored here: random variations in shape, size and intensity from pulse to pulse, lack 
of knowledge of the exact modulating function, the exact shape of the integrated 
profile, etc. Maybe there is some hope of making better prediction about the higher 
harmonics ($m > 3$) at higher frequencies ($> 100$ [cpp]), but DR2001 do not analyze 
these frequencies. The next section indicates why, if at all, the two spectra can 
probably be distinguished only at higher frequencies.

\subsection{Time domain discussion}

Let $t_n$ be the time of occurrence of the $n^{\mathrm{th}}$ pulse in a PPM signal that
is modulated by a pure tone, of phase $\phi_m$, and frequency $1 / P3$ [cpp], implying 
angular frequency $\omega_m = 2 \pi / (P3 P)$,
\begin{equation}
t_n + \tau \sin \left ( \omega_m t_n + \phi_m \right ) = n P,
\end{equation}
\noindent
where $\tau$ is the maximum position departure (see eq.~6-2-2 of SBS1966). Let $\Delta 
t_n$ be the difference between $t_n$ and $n P$, its time of occurrence in the un 
modulated case,
\begin{equation}
\Delta t_n = - \tau \sin \left ( \Phi_n + \omega_m \Delta t_n \right ),
\end{equation}
\noindent
where $\Phi_n  = 2 \pi n / P3 + \phi_m$. In \p0943, $\tau \approx 0.021 P$, and the 
original $P3 \approx 2.15$, so the phase $\omega_m \tau = 2 \pi \tau / (P3 P) \approx 
0.06$, which is a small quantity. Furthermore, $\Delta t_n \le \tau$; therefore,
\begin{equation}
\Delta t_n \approx - \tau \sin \left ( \Phi_n \right ) \left [ 1 - \tau \left \{ 
0.5 \omega_c - \Delta \omega_m \right \} \cos \left ( \Phi_n \right ) \right ] 
\end{equation}
\noindent
where $\omega_c = 2 \pi / P$ is the sampling frequency (angular) and $\Delta \omega_m
= 0.5 \omega_c - \omega_m$. In \p0943, $\Delta \omega_m$ ($= 2.862 - 2.662 = 0.2$) is 
much smaller than either $0.5 \omega_c$ or $\omega_m$.

Now consider a PPM signal that is modulated by a pure tone, of phase $\phi^\prime_m = 
-\phi_m$, and a frequency that is the alias of the earlier one; i.e., of angular 
frequency $\omega^\prime_m = \omega_c - \omega_m$. Let it have the same magnitude of 
drift but opposite in direction, i.e., $\tau^\prime = - \tau$; Then it can be shown 
that 
\begin{equation}
\Delta t^\prime_n \approx - \tau \sin \left ( \Phi_n \right ) \left [ 1 + \tau \left \{ 
0.5 \omega_c + \Delta \omega_m \right \} \cos \left ( \Phi_n \right ) \right ],
\end{equation}
\noindent
for integer $n$. For non-integer $n$, which implies that the pulses start at some 
arbitrary phase within the period, a constant phase adds to $\Phi_n$ in eq.~(4).

Now, the observed time series corresponds to either eq.~(3) or eq.~(4). If one knows 
its absolute phase, then one directly compares the observation to those two equations, 
and thus determines if the observed frequency is original or aliased. However, this is 
rarely the case. So, one has to distinguish between the the time series $\pm 0.25 
\omega_c \tau^2 \sin \left ( 2 \Phi_n \right )$ of eq.~(3) and eq.~(4). This is 
possible only if the sampling interval used for the observations is much smaller than 
the quantity $0.5 \omega_c \times \tau^2$, which is $\approx 1.52$ ms in \p0943. 
However the sampling interval of DR2001 (1.006 ms) is barely smaller than this.
Therefore it is probably impossible for DR1999 and DR2001 to distinguish between the 
original drift frequency and its alias with opposite drift direction. This argument 
also holds for saw-tooth modulation drifting pattern, which can be Fourier decomposed 
into harmonics of the fundamental modulating frequency $\omega_m$. 

Now one can understand the result of the previous section - a time resolution $<< 
1.52$ ms implies a minimum Nyquist frequency of $ >> 0.5 / 0.00152 \approx 329$ Hz 
or 361 [cpp]. Indeed, the spectra in fig.~\ref{fig1} differ most in the range 500 to 
900 [cpp], mainly in the higher harmonics $m >> 3$. For example, the 15$^{\mathsf{th}}$ 
harmonics (aliased) at frequency $679.975$ [cpp] differ in amplitude by $\approx$ 
107\%; and their average amplitude is about 24\% of the maximum amplitude of 
pulsar harmonics.

In summary, DR2001 can hope to distinguish between the original and aliased spectral
feature in \p0943 by predicting the two amplitude spectra at frequencies much larger 
than 100 [cpp], particularly for the harmonics $m \ge 3$ to $15$, and comparing those 
with observations. This they have not done.

\section{Aliasing claim by DR2001}

This section discusses the three main arguments that DR2001 appear to offer (their 
sections 3 and 5) regarding the spectral feature in \p0943.

\subsection{Amplitude spectrum argument}

In the longitude resolved spectrum in fig.~1 of DR2001, the narrow spectral feature 
falls at frequency 0.465 [cpp]. The spectrum of such a signal will naturally have a 
Nyquist sampling limit of 0.5 [cpp], by definition. To test whether this feature is 
aliased or not, one needs data with much faster sampling. DR2001 achieve this by 
obtaining the spectrum of the original time series (including zero padding) which is 
sampled at $1.006$ (ms). The corresponding spectrum certainly shows that the spectral 
feature is much stronger at 0.535 [cpp] than at 0.465 [cpp] (their fig.~4).

Therefore they argue (on the right side of page 442) that ``... the asymmetry 
between the 0.535 and 0.465 [cpp] features in fig.~4 is so great that we can regard 
them as a signature of an almost pure phase modulation''. This is justified, as seen
in fig.~\ref{fig1} above.  Then, after a short paragraph offering no new argument or 
evidence, they state that ``We can now be certain that the principal feature in fig.~1 
is the alias of a fluctuation the actual frequency which is greater than 0.5 [cpp] 
...''. This is unjustified.

\begin{figure}
\includegraphics[width=14.0cm]{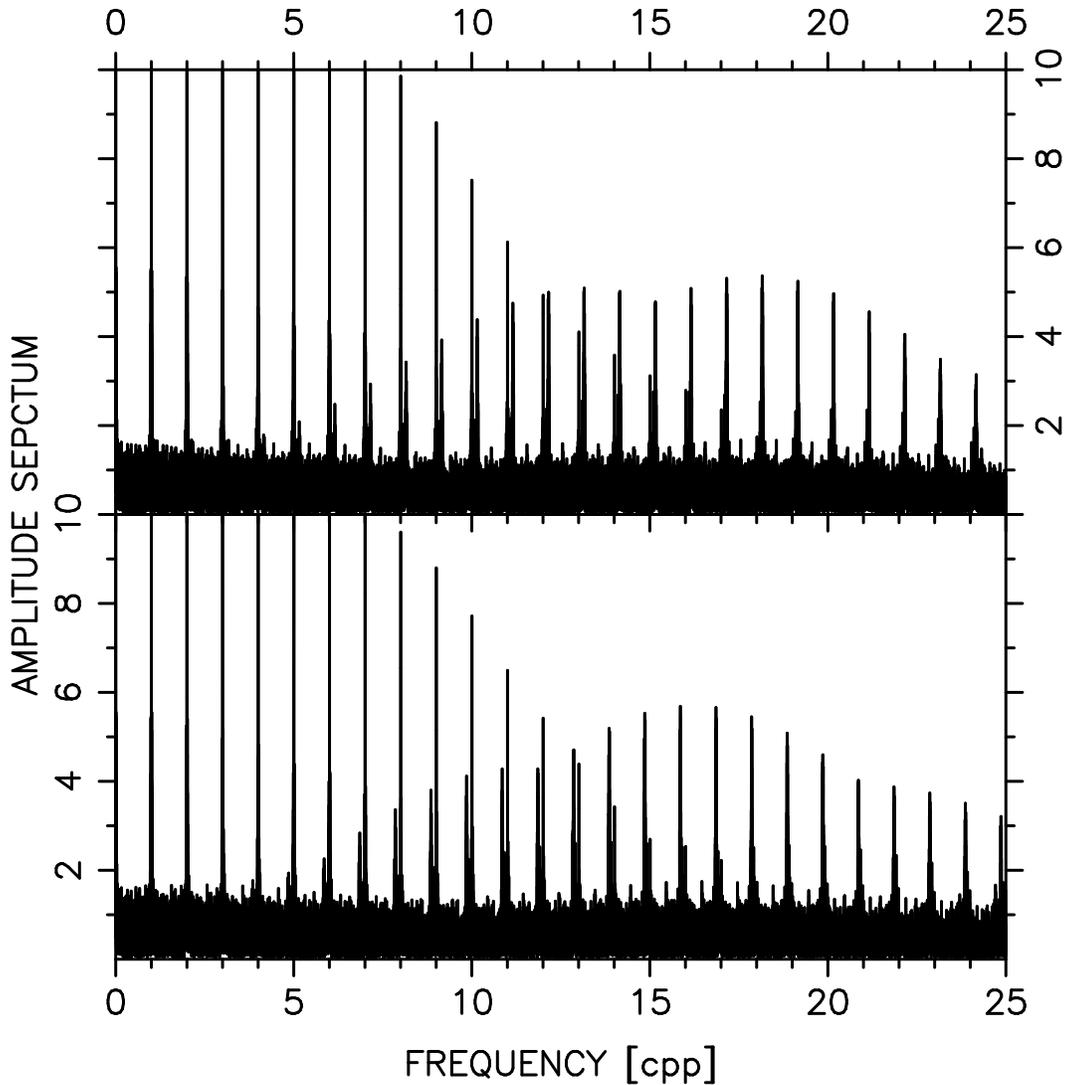}
\caption{
	{\bf Top panel}: Amplitude spectrum of 1024 periods of radio flux data of
	\q0031, obtained on 1996 March 31 using the ORT. The data was acquired in a 
	gated fashion and later zero padded., as in DR2001. The sampling interval was 
	6.5 ms and the pulsar has period 0.9430 (sec). Only a small range of frequency 
	has been shown for better visual comparison. {\bf Bottom panel}: The same as 
	in the top panel, except that the time sequence within each period is reversed, 
	while the period sequence itself is retained.
	}
\label{fig2}
\end{figure}

It is purely fortuitous that the aliased feature is stronger than the original in their 
fig.~4. It depends upon the specific combination of {\bf true} $P3$, {\bf true} drift 
direction and the {\bf exact} shape of the IP of \p0943. Fig.~\ref{fig2} verifies the 
above argument. It shows the amplitude spectra of \q0031, another drifting pulsar,
observed using Ooty Radio Telescope (ORT). The top panel corresponds to fig.~3 of DR2001. 
In the bottom panel of fig.~\ref{fig2} the data is altered: the period sequence is 
retained, but within each period the time sequence is reversed. This could have 
occurred either due to (1) \q0031 having the aliased drift frequency but the same drift 
direction, or (2) it having the original drift frequency but opposite drift direction. 
The asymmetry has also got reversed in the bottom panel of fig.~\ref{fig2}.

\subsection{Phase spectrum argument}

DR2001 do not use the phase information corresponding to their fig.~4, but use that 
available in the longitude resolved spectrum (their fig.~1). They compare the rate of 
change of phase, with longitude, of the 0.465 [cpp] spectral feature, with that of two 
other much weaker spectral features occurring at 0.071 [cpp] and 0.607 [cpp], which 
are supposed to be the aliased first and second harmonics of the fundamental. 
They claim on page 443 that ``the harmonicity of these phase rates argues strongly 
that the 0.071 [cpp] and 0.607 [cpp] features are indeed the aliases of the second 
and third harmonics of the primary feature ...'', which is justified. 

However, this will be true irrespective of whether the original frequency is 0.465 
[cpp] or 0.535 [cpp]; this is obvious from Fourier theory. The only difference is 
the inverted phase relation between the fundamental and its (aliased) harmonics, 
about which one has no independent information anyway. This author is confused about 
what DR2001 intended in invoking this discussion. This author is most confused about 
a sentence in the last para of their section 3, the section that is supposed to 
justify their claim of aliasing. It goes ``The ambiguity, however, between the 
remaining two possible combinations -- namely, $P3 < 2 P$ (thus implying negative 
drift) and $P3 > 2 P$ (positive drift) can not be fully resolved through the above 
analysis, ...''. Presumably this is what \cite{SO1975} implied, with which DR2001 
apparently disagreed earlier. It appears that either DR2001 are contradicting 
themselves, or this author has not understood what was it in \p0943 that was 
``otherwise multiply folded'' which DR2001 have managed to ``unfold'' (last para 
of their section 3).

\subsection{``Modulation on modulation'' argument}

\begin{figure}
\includegraphics[width=14.0cm]{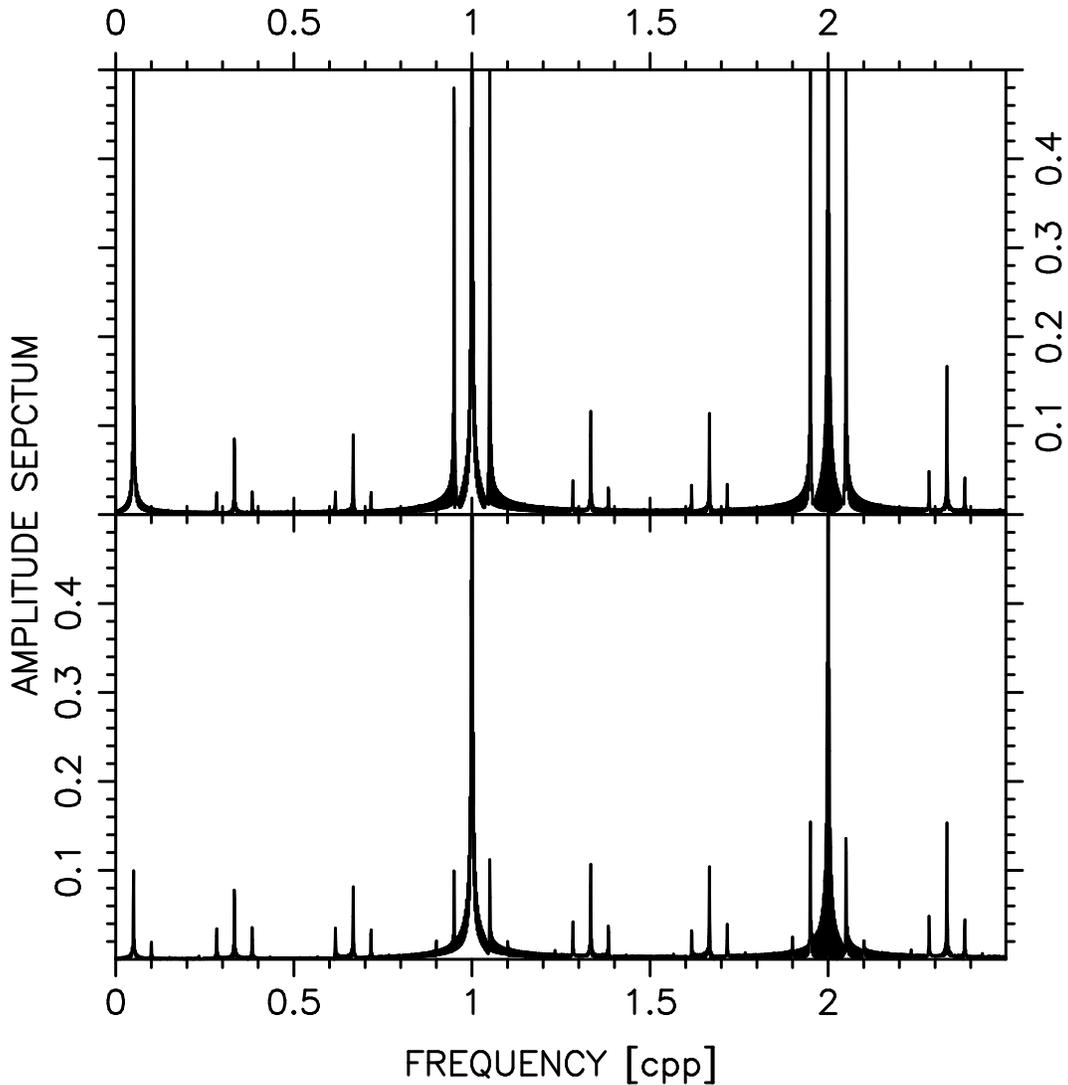}
\caption{
	Same as in fig.~\ref{fig1}, except (1) $1024 \times 1024$ samples used, (2)
	sampling interval of 1.0 ms, (3) modulating $P3 = 3$ periods. Only a small 
	range of frequency has been shown for better visual comparison. {\bf Top 
	panel}: The PPM signal is additionally amplitude modulated by a tone of 
	period 20 [P], and relative amplitude 0.5. {\bf Bottom panel}: The PPM 
	signal is additionally position modulated by a tone of period 20 [P], and 
	maximum time departure of 0.010 [P].
	}
\label{fig3}
\end{figure}

DR1999 claim on their page 1010 that the symmetrical sidebands associated with the 
primary feature in the fluctuation spectrum represent ``an amplitude modulation on
the phase modulation''. These sidebands fall 0.027 [cpp] higher and lower than the
primary feature. Since $0.535 / 0.027 \approx 20$, an integer value, DR2001 claim
in their conclusion that ``it is at this point that we have conclusive evidence
that the aliasing question is resolved ...''.

Fig.~\ref{fig3} is similar to fig.~\ref{fig1} except that (1) 4 times fewer samples 
and 4 times larger sampling interval, and (2) the drifting $P3 = 3$ periods, for 
better visibility. However, in the top panel the pulsar signal is additionally 
amplitude modulated by a sinusoid of frequency 0.05 [cpp]; of relative magnitude of 
modulation of 50\%, for the sidebands on the position modulating harmonics to show 
up significantly. The pulsar harmonics also have the symmetric sidebands, but of 
much larger amplitude. In the bottom panel of fig.~\ref{fig3} the above signal is 
instead additionally position modulated, by a sinusoid of the the same frequency, 
and a maximum time departure of 0.010 [P]. Now also the pulsar harmonics have the 
symmetric sidebands, but much smaller in amplitude.

From fig.~\ref{fig3} shows that both an additional amplitude modulation or an 
additional position modulation of the pulsar signal in fig.~\ref{fig1} gives
rise to sidebands around the drifting harmonics. However, in the former case the
side bands are much stronger around the pulsar harmonics (at 1 [cpp], 2 [cpp], 
etc.), and in the latter case they are much stronger around the drifting harmonics, 
which is what DR2001 observe. Therefore it is as likely, if not more likely, that 
DR2001 have noticed an additional position modulation over and above their drifting 
sub pulse pattern. DR2001 do not reason why they prefer the former over the latter.

The foundation of DR2001 vanishes if they are unable to justify the amplitude 
modulation of the drifting sub pulses.

\section{Summary}
DR2001's  claim of aliasing in \p0943 may rest purely upon an arbitrary assumption, 
that of a very steady ``amplitude modulation on the phase modulation''. Recent
claims \citep{ES2002} to have independently verified the results of DR2001 are,
in the opinion of this author, not relevant here because (1) they are model 
dependent, and (2) they work with the phase resolved spectrum, which is inadequate 
by definition to discuss the issue of aliasing.

This research has made use of NASA's Astrophysics Data System (ADS) Bibliographic 
Services.

\vfill
\eject
\end{document}